# Influence of High-Performance Image-to-Image Translation Networks on Clinical Visual Assessment and Outcome Prediction: Utilizing Ultrasound to MRI Translation in Prostate Cancer

Mohammad R. Salmanpour[1,2*] (*m.salmanpour@ubc.ca*), Amin Mousavi[3], Yixi Xu[2], William B Weeks[2], Ilker Hacihaliloglu[1,4]

[1]Department of Radiology, University of British Columbia, Vancouver, BC, Canada
[2]AI for Good Research Lab, Microsoft Corporation, Redmond, WA
[3]Department of Computer, Abhar Branch, Islamic Azad University, Abhar, Iran
[4]Department of Medicine, University of British Columbia, Vancouver, BC, Canada

(*) Author to whom correspondence should be addressed.

## Abstract

**Purpose:** Image-to-image (I2I) translation networks have emerged as promising tools for generating synthetic medical images; however, their clinical reliability and ability to preserve diagnostically relevant features remain underexplored. This study evaluates the performance of state-of-the-art 2D/3D I2I networks for converting ultrasound (US) images to synthetic MRI in prostate cancer (PCa) imaging. The novelty lies in combining radiomics, expert clinical evaluation, and classification performance to comprehensively benchmark these models for potential integration into real-world diagnostic workflows.
**Methods:** A dataset of 794 PCa patients was analyzed using ten leading I2I networks to synthesize MRI from US input. Radiomics feature (RF) analysis was performed using Spearman correlation to assess whether high-performing networks (SSIM>0.85) preserved quantitative imaging biomarkers. A qualitative evaluation by seven experienced physicians assessed the anatomical realism, presence of artifacts, and diagnostic interpretability of synthetic images. Additionally, classification tasks using synthetic images were conducted using two machine learning and one deep learning model to assess the practical diagnostic benefit.
**Results:** Among all networks, 2D-Pix2Pix achieved the highest SSIM (0.855±0.032). RF analysis showed that 76 out of 186 features were preserved post-translation, while the remainder were degraded or lost. Qualitative feedback revealed consistent issues with low-level feature preservation and artifact generation, particularly in lesion-rich regions. These evaluations were conducted to assess whether synthetic MRI retained clinically relevant patterns, supported expert interpretation, and improved diagnostic accuracy. Importantly, classification performance using synthetic MRI significantly exceeded that of US-based input, achieving average accuracy and AUC of ~ 0.93±0.05.
**Conclusion:** Although 2D-Pix2Pix showed the best overall performance in similarity and partial RF preservation, improvements are still required in lesion-level fidelity and artifact suppression. The combination of radiomics, qualitative, and classification analyses offered a holistic view of the current strengths and limitations of I2I models, supporting their potential in clinical applications pending further refinement and validation.

## 1. Introduction

Prostate cancer (PCa) is the second most common cancer and the fifth leading cause of death in men aged 45–60 worldwide [1]. Diagnosis methods include prostate-specific antigen (PSA) testing, magnetic resonance imaging (MRI), and ultrasound (US), with MRI providing high-resolution imaging [2] but limited by its high cost, and US offering a cost-effective, real-time alternative with lower sensitivity and specificity [3]. Deep learning (DL) has improved MRI's diagnostic accuracy and addressed limitations of US [4]. Recent research focuses on Image-to-Image (I2I) translation networks, which use convolutional neural networks (CNNs) and generative adversarial networks (GANs) to convert US into high-quality images like MRI, computed tomography (CT), or X-ray [5]. Promising studies include self-supervised methods for fetal brain MRI synthesis [6], stacked GANs for pseudo-CT generation [7], pseudo-anatomical displays from US data [8], generative attention networks for spine synthesis [9], real-time volumetric registration for surgical guidance [10], and hierarchical variational auto-encoders for generating MRIs from incomplete US data [11].

Synthesized image quality is typically evaluated using metrics like Mean Absolute Error (MAE), Mean Square Error (MSE), Structural Similarity Index (SSIM), and Peak Signal-to-Noise Ratio (PSNR) [12]. However, these metrics may not fully capture biological complexity [13, 14, 15, 16]. Some studies assess improvements in downstream tasks like classification or segmentation [17]. Radiomic features (RF), such as spatial distribution, shape, intensity, and texture, extracted from synthetic images, provide additional insights, ensuring critical diagnostic information is retained during translation [18].

RFs, enhanced by imaging and AI advancements, show promise in predicting clinical outcomes in PCa and supporting personalized treatments [19]. [20] introduced an open-source framework for RF workflows to classify PCa using the University of California-Los Angeles PCa Index (UCLA). AI and RFs have been explored for identifying predictive and prognostic biomarkers [21], combining MRI-based RFs with machine learning to predict Extraprostatic Extension in high-risk PCa patients [22], and using RFs from pre-treatment CT scans to predict 5-year progression-free survival [23]. Integrated RF models, merging MRI-derived RFs with clinical features, have predicted pelvic lymph node invasion [24], while MRI-derived RF biomarkers have been used non-invasively to predict PCa grade and surgery suitability, guided by clinical expertise [25].

A recent study [26] showed that RF models, which extract detailed features from medical images, outperformed DL models in predicting PCa outcomes by quantifying tumor characteristics. However, DL-based models have been widely explored [27]. Examples include the RegNetY-320 model for prediction of UCLA scores [28], texture-based DL for detecting significant PCa [29], deep residual CNNs for classification [30], and architectures like long short-term memory and ResNet-101 for direct outcome prediction [31]. Transfer learning CNNs [32], NASNetLarge for whole-slide images [33], dual CNNs for nuclei detection and classification [34], and automated CNN pipelines for diffusion-weighted imaging [35] have also been used. Yet, RF analysis in the context of image synthesis remains unexplored.

This study compares RFs from high-resolution and synthesized MRI data for PCa, using 10 DL I2I networks to generate prostate MRI from US, a first for detecting cancerous lesions. It evaluates whether high-performing networks (SSIM>85%) capture low-level RFs, linking image quality to clinical feature detection. Insights from seven clinicians and diagnostic tests reveal the potential of synthetic MRI to enhance PCa management.

## 2. Materials and Methods

### 2.1. Patient Data and Preprocessing Steps

We utilized data from 794 patients with PCa sourced from The Cancer Imaging Archive (TCIA) [36], all of whom underwent 3D transrectal US, T2-weighted MRI, and had corresponding segmentation masks on both modalities. This dataset originates from biopsy sessions using the Artemis biopsy system, which integrates both targeted and systematic sampling. Targeted biopsies were guided through nonrigid registration (fusion) between preoperative MRI and real-time US, enabling accurate sampling of MRI-defined regions of interest, while systematic biopsies followed a standard 12-core digital template. Core locations were recorded by the Artemis system using encoder kinematics of a mechanical arm and registered relative to the US volume; MRI coordinates of the biopsy cores were also documented. Multiparametric MRI, including T2-weighted, diffusion-weighted, and perfusion-weighted sequences, was used to define targets, though only T2-weighted data are included in this dataset. Lesions were scored using the UCLA scoring system, introduced in 2010 as a structured framework aligned with ESUR PI-RADS and later refined in accordance with PI-RADS version 2 [37]. The system assigns scores from 1 (very low suspicion) to 5 (very high suspicion) based on imaging characteristics [38]. For the purposes of this study, scores 1–3 were classified as low-risk, while scores 4–5 were categorized as high-risk. The UCLA index has demonstrated reliable psychometric validity, particularly among older male patients, and captures both general and condition-specific quality-of-life concerns [39]. Table 1 shows the baseline characteristics of the patients included in the study.

MRI scans were acquired on Siemens 3T Trio, Verio, or Skyra scanners (Erlangen, Germany), using a transabdominal phased-array coil in all cases and an endorectal coil in a subset. The primary protocol was a 3D T2-weighted SPACE sequence (TR/TE = 2200/203 ms, matrix = 256 × 205, FOV = 14 × 14 cm, slice thickness = 1.5 mm), with some cases using 3D T2-weighted TSE (TR/TE = 3800–5040/101 ms) or external variants. The average in-plane voxel size was ~0.55 × 0.68 mm. US scans were performed with a Hitachi Hi-Vision 5500 (7.5 MHz) or Noblus C41V (2–10 MHz) end-fire probe, rotating 200 degrees to generate isotropic 3D volumes. All original DICOM files preserved raw intensity values. During preprocessing, US and MRI images were aligned by clinical collaborators, then cropped around the prostate center and resampled to a standardized size of 128 × 128 × 64 mm³. Min–max normalization was applied to each volume individually to harmonize intensity distributions across patients and imaging modalities. Patients were enrolled consecutively from the UCLA Clark Urology Center based on elevated PSA or abnormal imaging findings, and all underwent standard-of-care biopsy. Importantly, the dataset includes private DICOM metadata (e.g., tag (1129, "Eigen, Inc", 1016) for voxel size), which is essential for rendering anatomy and STL surfaces, especially in multi-frame US data.

**Table 1.** Summary of patient characteristics included in the study. mm: millimeter, ng/mL: nanograms per milliliter, NA: Data not available.

| Total number of patients | 794 |
|---|---|
| Age | 66 ± 5 years |
| PSA | 9.6 ± 11.38 ng/mL |
| Cancer Length | 3 ± 2.75 mm |
| % Cancer in Core | 24.62 ± 23.47 mm |
| Core Fragment #1 Tissue Length | 14.53 ± 5.15 mm |
| Core Fragment #2 Tissue Length | 3.08 ± 2.56 mm |
| Core Fragment #3 Tissue Length | 1.79 ± 1 mm |
| UCLA Score | Score I (4%), Score II (2%), Score III (38%), Score IV (33%), Score V (23%) |
| Primary Gleason | Score III (67%), Score IV (7%), Score V (1%), NA (25%) |
| Secondary Gleason | Score III (54%), Score IV (18%), Score V (2%), NA (26%) |
| Total Gleason | Score VI (50%), Score VII (21%), Score VIII (2%), Score IX (2%), Score X (<1%), NA (25%) |

## 2.2. DL-Based I2I Networks

This Study investigates ten I2I networks including 2D-CycleGAN, 2D-Pix2Pix, 2D-DiscoGAN, 2D-GcGAN, 2D-DualGAN, 2D-ContourDiff, 3D-CycleGAN, 3D-AutoEncoder, 3D-UNET, 3D-Med-DDPM to synthesize MRI from US images [40, 41, 42]. All 2D networks were trained using each 2D image from the volumetric data. The dataset, comprising 794 patients, was split into three sections: 75% for training, 10% for validation (and model selection), and 15% for external testing. The networks' performance was evaluated on 3D volumetric data using four metrics: MAE and MSE—both computed on images normalized to the [0, 1] intensity range before training and evaluation—as well as SSIM and PSNR. We performed 5-fold cross-validation and reported an average across all the experiments. Individual 2D slices from the 3D volumetric data were used as input for the 2D models, and the 3D volume was reconstructed by integrating these slices. Supplemental Table S1 shows different parameters utilized to tune I2I networks.

## 2.3. RF Analysis

We utilized RF generator within ViSERA [43], extensively standardized in reference to the Image Biomarker Standardization Initiative to extract a total of 186 standardized RFs from the segmented prostate gland, including 2 local intensity (LI), 18 intensity-based statistics (IS), 23 intensity histogram (IH), 7 Intensity-Volume Histogram (IVH), and 136 texture features containing gray-level co-occurrence matrix (GLCM; 50 features), gray-level run-length matrix (GLRLM; 32 features), gray-level size zones (GLSZM; 16 features), gray-level distance zone matrix (GLDZM; 16 features), neighborhood gray-tone difference matrix (NGTDM, 5 features), and neighboring gray-level dependence matrix (NGLDM; 17 features). RF analysis was conducted using the Spearman correlation function and paired t-test. Additionally, morphological characteristics were not considered in this study because identical masks were used to extract these features from different images, including both original and synthetic MRIs. Complete names and specifics of each RF sub-category and codes are detailed in supplemental Table S2.

## 2.4. Qualitative Analysis

Medical professionals' evaluation of synthetic images is crucial for ensuring their clinical accuracy and reliability. Through detailed visual inspections, experts validate diagnostic relevance, refine image generation algorithms, and enhance their utility in diagnostics, training, and research. This collaboration between technology and medicine ensures the clinical efficacy of synthetic imaging, paving the way for improved patient outcomes and more efficient healthcare delivery. For qualitative validation, seven physicians with over 5 years of MRI and US experience evaluated synthetic MRI images, distinguishing them from originals and answering eight comparison questions (Table 1, rows 2–9).

## 2.4. Classification Analysis

This study applied two methods: RF-based machine learning and DL-based frameworks. In the RF approach, real and translated MRI and US datasets were normalized using a min-max function and classified with a Random-Forest (RandF) algorithm. Dimensionality reduction via Principal Component Analysis (PCA) was employed to streamline processes, reduce complexity, and enhance performance [44]. For the DL framework, the ResNet50 architecture was used, leveraging its deep structure to analyze normalized input images and accurately classify PCa patients into high- and low-risk categories [45]. Supplemental Table S3 outlines experiments using real and synthetic MRI for PCa classification, dividing 794 patients into training (75%)/validation (10%)/testing (15%) datasets as per the I2I methodology. Metrics like average accuracy and Area Under the Curve (AUC) were evaluated with 5-fold cross-validation. Supplemental Section 1.2 shows different parameters utilized to tune classifiers.

## 3. Results

### 3.1. DL-Based Image Translation Quantitative Assessment

As shown in Figure 1, the 2D-Pix2Pix network significantly outperformed all other generative models across all evaluation metrics—MAE, MSE, SSIM, and PSNR—with statistically significant differences ($P < 0.01$, paired $t$-test). It achieved an average MAE of $0.026 \pm 0.007$, MSE of $0.001 \pm 0.001$, SSIM of $0.855 \pm 0.032$, and PSNR of $28.831 \pm 2.067$, indicating superior reconstruction accuracy and structural consistency. Among the remaining models, diffusion-based networks (2D-ContourDiff and 3D-Med-DDPM) ranked second overall, delivering competitive performance. In contrast, CycleGAN-based models, particularly the 3D variant, showed the lowest accuracy and poorest similarity scores. 3D-AutoEncoder, 3D-UNET, and 2D-GAN variants such as DualGAN and DiscoGAN demonstrated intermediate performance. These results emphasize the effectiveness of supervised paired-image translation methods like Pix2Pix, while also highlighting the growing promise of diffusion models in medical image synthesis. Other networks, including 2D-GcGAN and 2D-CycleGAN, performed below the level of 3D-UNET, indicating limited suitability for high-fidelity image translation.

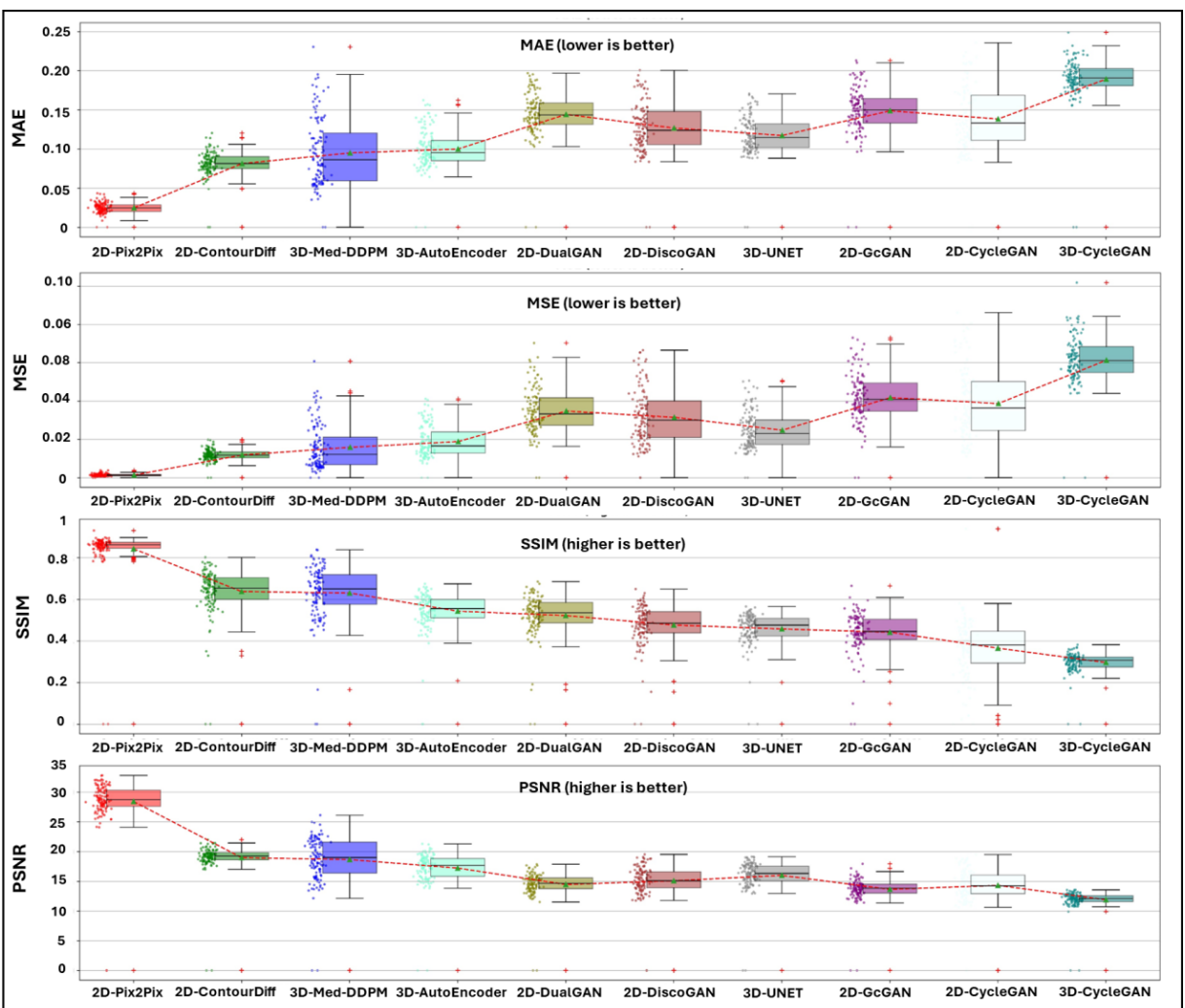

**Fig. 1.** A distribution of 4 quantitative evaluation metrics: MAE, MSE, PSNR, and SSIM for ten generative networks in synthesizing MRI images from ultrasound images.

Figure 2 shows qualitative results from 4 external testing examples of synthetic MRI images provided by 2D-Pix2Pix algorithm. The figure includes US, original MRI, synthetic MRI, and the difference between original and synthetic MRI for 4 patients. In Figure 2, Patient 3 presents a compelling example of the challenges associated with lesion fidelity in synthetic MRI. Notably, a hyperintense region is observed in the synthetic image (orange circle) that does not correspond to any suspicious area in the original MRI. This artifact, likely introduced during the image synthesis process, could lead to a false-positive interpretation in a clinical setting. Conversely, a true lesion with high intensity, clearly visible in the original MRI (green circle), is not accurately preserved in the synthetic image, raising concerns about false-negative outcomes. These discrepancies underscore the importance of rigorous validation and comprehensive evaluation—both qualitative and quantitative—to assess the diagnostic reliability of synthetic MRI. This specific case highlights the limitations of current generative models and the need for further refinement to ensure the accurate preservation of clinically relevant features. Supplemental Figure S1 illustrates four examples of synthetic MRI images generated by each of the other nine networks.

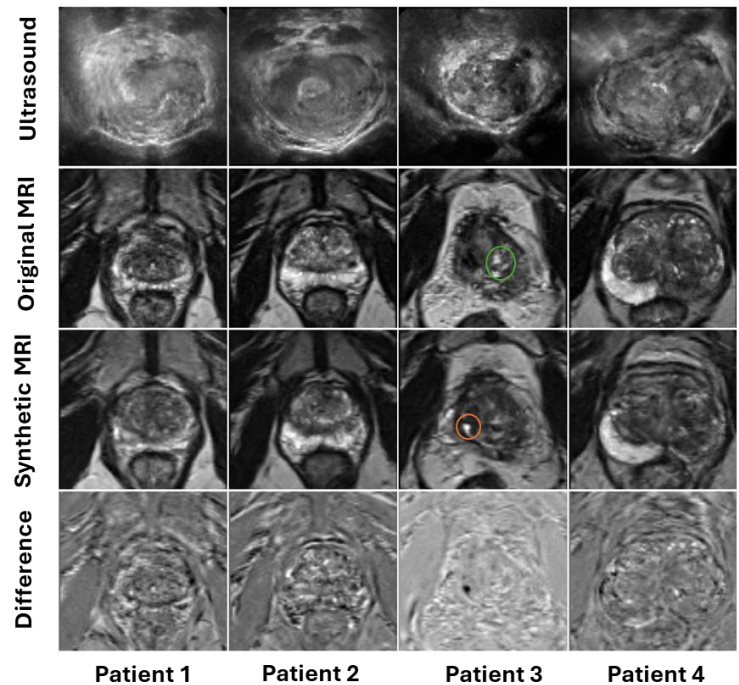

**Fig. 2.** Four examples of synthetic MRI images provided by 2D-Pix2Pix with SSIMs>0.85. Rows show Ultrasound, Original MRI, Synthetic MRI, difference between original and synthetic MRI images. For Patient 3, the orange circle highlights a false-positive hyperintensity introduced in the synthetic MRI, while the green circle marks a true lesion present in the original MRI that was not preserved in the synthetic version.

## 3.2. Qualitative Analysis

Seven experienced physicians qualitatively evaluated synthetic MRI images generated by the Pix2Pix network, assessing anatomical fidelity, tissue contrast, boundary delineation, and the presence of artifacts. While all could distinguish synthetic from original MRI despite an average SSIM > 0.85, they agreed the synthetic images lacked the detailed quality of the originals (Q1). Practitioners noted artifacts as key markers distinguishing synthetic MRI from originals (Q2) and found diagnosis with synthetic MRI, even at SSIM > 0.85, challenging compared to original MRI (Q3). While synthetic MRI added no diagnostic value over original MRI (Q5), some experts recognized its potential over US imaging (Q6). Despite challenges with resolution and contrast discrepancies (Q7), experts highlighted the lack of detailed anatomical information in synthetic MRI but acknowledged its potential clinical benefits if the detail improves, supporting its integration into practice (Q8 and Q9).

Since we encountered two critical scenarios—(i) the omission of true lesions in synthetic MRI, and (ii) the introduction of false-positive hyperintensities not present in the original MRI—relying solely on qualitative evaluation was not sufficient. While the qualitative assessment offered valuable clinical insight into image quality, diagnostic usability, and artifact perception, it remained inherently subjective and could not fully quantify the discrepancies in lesion representation. Moreover, in cases where synthetic images introduced suspicious regions with no reference in the original MRI, conventional similarity metrics were ineffective due to the absence of a ground truth for comparison. Following this realization and based on feedback from our multidisciplinary team—including imaging experts, physicians, and medical physicists—healthcare professionals recommended conducting a deeper, more objective analysis. RF analysis was identified as a robust approach to systematically investigate such discrepancies. Therefore, we expanded our study to include a comprehensive RF-based evaluation, as detailed in the following section. RFs capture fine-grained, quantifiable characteristics of medical images such as intensity, texture, and shape—attributes that might be visible or not to the human eye. This enabled us to assess whether diagnostically relevant features were faithfully preserved or distorted in the synthetic images. By applying RF analysis across both challenging scenarios, we were able to identify systematic patterns of agreement and deviation between the synthetic and original MRI images. Ultimately, integrating this feature-level, standardized evaluation with our earlier qualitative review provided a more complete and clinically meaningful assessment of the fidelity and limitations of synthetic MRI.

**Table 2.** Qualitative analysis of synthetic MRI images evaluated by seven medical doctors (D). The images were generated using the Pix2Pix network.

| Questions (**Q**), Scoring system: **0**=zero, **1**=low, **2**=intermediate, **3**=high, **4**=very high | **D1** | **D2** | **D3** | **D4** | **D5** | **D6** | **D7** |
|---|---|---|---|---|---|---|---|
| **Q1:** After specifying synthetic and original MRIs for you, how would you rate the overall quality of synthetic MRI images compared to the original MRI? (score: higher, better) | 1 | 2 | 1 | 1 | 1 | 1 | 1 |
| **Q2:** Are there any noticeable artifacts or inaccuracies in the synthetic MRI images? (score: higher, worse) | 4 | 2 | 4 | 4 | 3 | 3 | 3 |
| **Q3:** How confident are you in making a diagnosis based on synthetic MRI images versus original MRI? (score: higher, better) | 1 | 1 | 1 | 1 | 1 | 1 | 1 |
| **Q4:** Do synthetic MRI images offer any additional diagnostic information compared to the original MRI images? How much? (score: higher, better) | 0 | 0 | 0 | 0 | 0 | 0 | 0 |
| **Q5:** Do synthetic MRI images offer any additional diagnostic information compared to the original Ultrasound images? How much? (score: higher, better) | 2 | 2 | 3 | 2 | 3 | 3 | 2 |
| **Q6:** How do you assess the resolution and contrast of the synthetic MRI images, compared to the original MRI images? (score: higher, better) | 1 | 2 | 1 | 2 | 2 | 1 | 2 |
| **Q7:** In your opinion, how much are the potential clinical benefits of using synthetic MRI images? (score: higher, better) | 4 | 3 | 3 | 4 | 3 | 4 | 3 |
| **Q8:** Would you support the integration of synthetic MRI technology into regular clinical practice? How much? (score: higher, better) | 4 | 4 | 4 | 4 | 4 | 4 | 4 |

### 3.3. RF Analysis

Koo and Li [46] classified correlation coefficients as poor (< 0.50), moderate (0.50–0.75), good (0.75–0.90), and excellent (> 0.90), with a threshold of 0.50 employed in this study to distinguish between groups. Based on this analysis, our RF evaluation, presented in Supplemental Table S4, demonstrates that RFs extracted from synthetic MRI can be grouped into three distinct categories according to their correlation with the corresponding RFs from the original MRI. Specifically, the table quantifies the similarity between each RF obtained from the synthetic images and its counterpart from the original MRI using correlation coefficients. This stratification reveals that some RFs retain a high degree of similarity and are strongly correlated, indicating robust preservation through the synthesis process. Others show moderate or low correlation, suggesting variability in how well different types of features are transferred from the original to synthetic domains. This categorization helps to better understand the fidelity of the synthetic MRI in preserving meaningful imaging biomarkers for downstream analysis.

Figure 3 shows the average absolute correlation coefficients of RFs between synthetic MRI and original MRI, highlighting that Pix2Pix networks enable better discovery of low-level RFs compared to other networks. Group 1 included 18 low-level RFs (1 IS, 2 NGLDM, 4 GLRLM, 2 GLSZM, 6 GLDZM, and 3 NGTDM) identified using synthetic MRI images generated by Pix2Pix (see Supplemental Table S4), with an average correlation coefficient of $0.745 \pm 0.119$ (see Figure 3). Notably, other networks, including 3D-UNET, 2D-GcGAN, and 2D-CycleGAN, also achieved an average correlation coefficient exceeding 0.700 for Group 1 RFs, despite their lower SSIM performance. Group 2 consisted of 76 RFs (5 IS, 17 IH, 2 IVH, 26 GLCM, 6 NGLDM, 12 GLRLM, 3 GLSZM, 3 GLDZM, and 1 NGTDM) identified exclusively from synthetic MRI images produced by the high-performance 2D-Pix2Pix algorithm (SSIM > 0.85). A proportional relationship between network performance and feature discovery was observed, with an average correlation coefficient of $0.598 \pm 0.078$. Group 3 comprised 93 RFs (2 LI, 12 IS, 6 IH, 5 IVH, 24 GLCM, 9 NGLDM, 16 GLRLM, 11 GLSZM, 7 GLDZM, and 1 NGTDM) that remained undetectable by any network, including the high-performance 2D-Pix2Pix, which demonstrated an average correlation coefficient of $0.307 \pm 0.118$.

|  | **Group 1** | **Group 2** | **Group 3** |
|---|---|---|---|
| **Ultrasound** | 0.653±0.126 | 0.126±0.111 | 0.134±0.097 |
| **2D-Pix2Pix** | 0.745±0.119 | 0.598±0.078 | 0.307±0.118 |
| **3D-AutoEncode** | 0.466±0.288 | 0.099±0.118 | 0.098±0.061 |
| **2D-DualGAN** | 0.665±0.122 | 0.166±0.080 | 0.101±0.075 |
| **2D-DiscoGAN** | 0.707±0.155 | 0.100±0.114 | 0.058±0.57 |
| **3D-UNET** | 0.700±0.129 | 0.153±.089 | 0.086±0.68 |
| **2D-GcGAN** | 0.728±0.128 | 0.126±0.124 | 0.114±0.64 |
| **2D-CycleGAN** | 0.733±0.125 | 0.150±0.120 | 0.064±0.64 |
| **3D-CycleGAN** | 0.601±0.203 | 0.120±0.084 | 0.075±0.62 |
| **2D-ContourDiff** | 0.714±0.128 | 0.297±0.125 | 0.187±0.147 |
| **3D-Med-DDPM** | 0.535±0.173 | 0.107±0.064 | 0.074±0.065 |

**Figure 3.** Average ± Standard Deviation of correlation coefficients of radiomic features between synthetic and original MRI.

### 3.4 Classification Analysis

In the RF framework (Figure 4, row 1), C1 (RFs from real MRI) achieved the highest accuracy (0.95 ± 0.05) and AUC (0.94 ± 0.05) with PCA + RandF, while C2 (RFs from US) yielded an accuracy of 0.88 ± 0.04 and AUC of 0.87 ± 0.04 using the same approach. C3, C6, and C11 (RFs from 2D-Pix2Pix, 2D-GcGAN, and 3D-UNET images) with PCA + RandF achieved an average accuracy and AUC of about 0.93, while other combinations (C4-C5, C7, C9, C10) with synthetic imaging for both training and testing showed metrics between those of RFs from real US and MRI data. As shown in row 2, without PCA, C9 (RFs from 3D-CycleGAN images) + RandF achieved the accuracy (0.64 ± 0.04) and AUC (0.62 ± 0.04), while C1 (RFs from real MRI) + RandF yielded an accuracy of 0.60 ± 0.04 and AUC of 0.58 ± 0.04. As shown in row 3, C1 (real MRI with ResNet50) achieved an accuracy of 0.63 ± 0.06 and AUC of 0.50 ± 0.04, while C8 and C9 (2D-ContourDiff and 3D-CycleGAN synthetic MRI) achieved a slightly higher accuracy of 0.64. Overall, the RF framework significantly outperformed the DL framework (P < 0.05, paired t-test). Moreover, Figure 5 presents Receiver Operating Characteristic (ROC) curves for various classification models, including traditional machine learning algorithms and the deep learning model across different input combinations as defined in Figure 4, enabling a clear visual comparison of their performance across a range of thresholds.

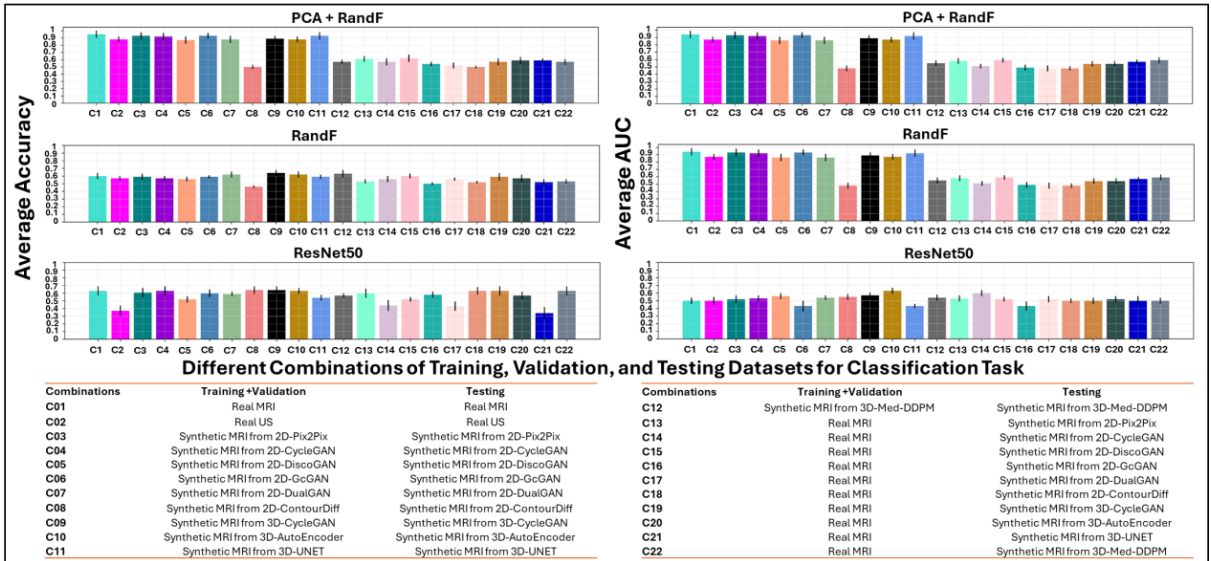

**Fig. 4.** Graphical representation of prediction accuracy and AUC metrics for radiomic and deep learning frameworks.

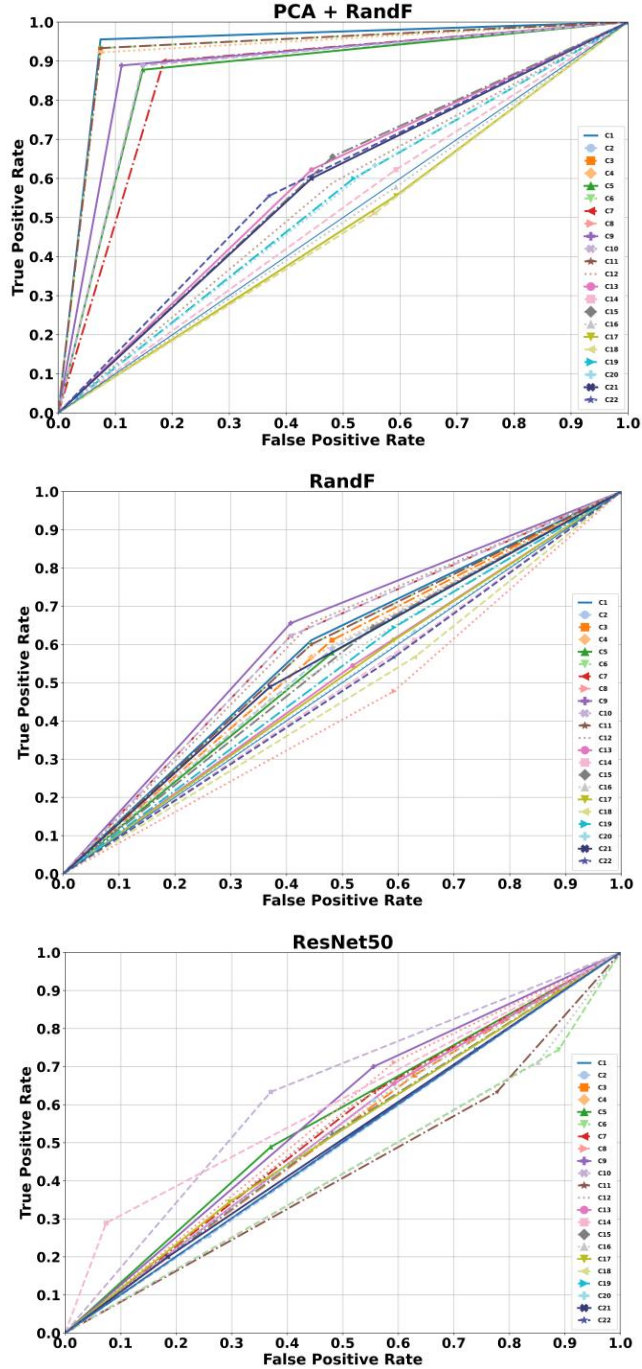

**Fig. 5.** ROC curves for the deep-learning and radiomics frameworks (C1–C22; definitions provided in Fig. 4).

### 4. Discussion

This study demonstrated that 2D-Pix2Pix networks achieved high-performance US-to-MRI generation, attaining an SSIM > 0.85 in terms of overall error and similarity, significantly surpassing the performance of nine other networks. In contrast, cutting-edge diffusion-based 2D-ContourDiff and 3D-Med-DDPM, which are designed to preserve anatomical structures using pixel-level constraints inspired by prior works in spatially-conditioned learning [41, 42], failed to generate high-quality images capable of accurately capturing the low-level features essential for medical professionals in decision-making and diagnosis. This study's RF analysis reveals three categories of RFs in US-to-MRI translation: (i) low-level RFs detectable by most networks, even low-performance ones, (ii) low-level RFs detectable only by high-performance networks, and (iii) low-level RFs currently undetectable by any existing networks. This detailed analysis, beyond traditional metrics like overall error and similarity, highlights limitations in current I2I networks to discover some RFs in group 3.

Traditional metrics like MAE, MSE, SSIM, and PSNR assess global similarity but often miss intricate low-level features crucial for medical imaging [13, 14, 15, 16]. While MAE/MSE measure pixel-wise errors, SSIM focuses on structural alignment, and PSNR evaluates noise, all fall short in capturing diagnostically significant details. RFs provide a superior alternative by offering quantitative insights into patterns, textures, and structures, enabling more precise evaluations of diagnostically relevant features in I2I translations. Supporting this, a recent study [47] demonstrated the clinical relevance of RFs by integrating them with PI-RADS scoring, demonstrating their ability to represent key clinical assessment features. PI-RADS scores 1 and 2 typically indicate normal or benign findings. PI-RADS 1 lesions, characterized by homogeneous signal intensity and spherical shapes on T2WI, align with RFs capturing high homogeneity and compactness, confirming their benign nature. PI-RADS 2 lesions, mildly hypo-intense on T2WI, appear as encapsulated nodules or wedge shapes; RFs quantify their mild intensity variations and geometry, indicating low cancer risk. PI-RADS score 3 represents an equivocal category with uncertain malignancy likelihood. Lesions exhibit mild hypo-intensity and complex shapes, such as wedges, with RFs capturing heterogeneity (entropy, intensity variation) and irregular shapes to quantify these ambiguous features. PI-RADS scores 4 and 5 indicate higher malignancy risk: PI-RADS 4 involves lesions < 1.5 cm with moderate hypo-intensity and irregular shapes, while PI-RADS 5 includes larger or invasive lesions. RFs quantify size, shape, and intensity variations to support these assessments.

RF analysis showed 74% of IH features in Group 2, reflecting network dependence on features that quantify tumor heterogeneity, microenvironment, and treatment responses. Metrics like Mean, Variance, Skewness, and Kurtosis analyze distribution patterns, while Median, Mode, and Percentiles assess central tendencies and variability. Entropy, Uniformity, and Gradient evaluate texture and edges, aiding tissue assessment and clinical decisions. GLCM features, used for texture analysis, were 53% in Group 2 and 47% in Group 3, highlighting network limitations in detecting homogeneity, contrast, and texture, crucial for distinguishing healthy and abnormal tissues. Most NGTDM features in Group 1 consistently correlated with network performance, analyzing gray-tone variations crucial for distinguishing malignant from benign lesions. LI and most IS features, primarily in Group 3, assist in detecting subtle pixel intensity changes within tumors. IVH features provide detailed intensity distribution insights, enhancing understanding of tumor diversity. Most GLRLM, GLSZM, NGLDM, and GLDZM features, found in Group 3, offer essential insights into tumor heterogeneity, spatial patterns, and microstructure, improving clinical decisions and patient care.

Qualitative analysis showed that differences in quality, artifacts, resolution, and contrast enabled physicians to distinguish synthetic from original MRI images. Despite SSIM values over 0.85, diagnosing synthetic images was more challenging due to missing low-level features critical for accuracy. While synthetic MRI images offered some diagnostic value compared to US images, enhancing their detail could improve clinical utility, especially in areas with limited MRI access.

The RF framework outperformed the DL framework, achieving an average accuracy of 0.95 using real MRI images with PCA and RandF, while DL accuracy of 0.64 came from 3D-CycleGAN-generated images with ResNet50. RF excelled due to PCA's dimensionality reduction and better generalization with real data, whereas DL models like ResNet50 struggled with limited datasets and the complexity of learning features effectively. Analyzing a network trained on real MRI images and tested with synthetic images showed no improvement. However, using consistent synthetic data for training and testing with PCA and RandF eliminated the domain gap, reduced noise, and improved classification accuracy compared to US.

This study is limited by variability in I2I network performance across datasets, including the small size of ours, which affects generalizability. Classifiers may not fully capture complex image features, and the lack of diverse datasets restricts insights. A recent study [48] improved MRI-to-CT conversion using a GLCM-based loss function to enhance texture quality, a method that could be adapted to our networks to improve RF discoveries in future work. Additionally, discrepancies such as the omission of true lesions or the introduction of false-positive artifacts in synthetic images, as discussed in the result subsection and illustrated in Figure 2, highlight the need for improved lesion-level fidelity and stronger constraints on anatomical realism during image synthesis.

## 5. Conclusion

This study finds that while 2D-Pix2Pix (SSIM > 0.85) improves MRI synthesis and RF recognition, significant advancements are needed to capture nuanced low-level features. Designing I2I networks to detect such features while balancing error and similarity index is essential. Integrating I2I networks with PCA and RandF achieved better results (accuracy of 0.93) than using US directly, though real MRI data performed slightly better (accuracy of 0.95).

**Acknowledgments.** This work was supported by the Mitacs Accelerate program grant number AWD-024298-IT33280 and Microsoft's AI for Good Lab. We acknowledge the support of the Canadian Foundation for Innovation-John R. Evans Leaders Fund (CFI-JELF) program [AWD-023869 CFI]. We acknowledge the support of the Natural Sciences and Engineering Research Council of Canada (NSERC), [AWD-024385]. Cette recherche a été financée par le Conseil de recherches en sciences naturelles et en génie du Canada (CRSNG), [RGPIN-2023-0357].

**Code and Data Availability.** Codes are publicly shared at: *https://github.com/MohammadRSalmanpour/Image-to-Image-translation*

**Conflict of Interest:** Mohammad R. Salmanpour, Yixi Xu, and William B. Weeks are affiliated with Microsoft Corporation. All other co-authors declare no conflicts of interest.